\documentclass[11pt,letterpaper,pdf]{article}
\usepackage{jheppub}
\usepackage{amsmath}
\usepackage{slashed}

\title{Two-loop Bhabha Scattering at High Energy beyond Leading Power
Approximation}

\author[a,b]{Alexander A. Penin\,}
\author[c]{and Nikolai Zerf\,}

\affiliation[a]{Department of Physics, University of Alberta,\\
  Edmonton AB T6G 2J1, Canada}
\affiliation[b]{Institut f{\"u}r Theoretische Teilchenphysik, Karlsruhe
  Institute of Technology (KIT),\\
  76128 Karlsruhe, Germany}
  \affiliation[c]{Institut f{\"u}r Theoretische Physik, Universit\"at Heidelberg,\\
D-69120 Heidelberg, Deutschland
}

\emailAdd{penin@ualberta.ca}
\emailAdd{zerf@ualberta.ca}

\abstract{We evaluate the two-loop ${\cal O}(m_e^2/s)$ contribution to the
wide-angle high-energy electron-positron scattering in the double-logarithmic
approximation. The origin and the general structure of the power-suppressed
double logarithmic corrections is discussed in detail.}


\begin{document}

\preprint{ALBERTA-THY-04-16, TTP16-021}

\maketitle


\section{Introduction}
\label{sec::int}
High-energy electron-positron or {\it Bhabha} scattering \cite{Bhabha:1936zz} is
among the classical applications of the perturbative  quantum  electrodynamics
(QED). Beside its phenomenological importance as a standard candle for
luminosity calibration at the electron-positron colliders, Bhabha scattering has
become a testing ground for the new techniques of the multiloop calculations.
The analysis of high-order corrections to this process often sheds new light on
perturbative structure of gauge theories. In general the radiative corrections
for the scattering of two massive particles are known  only in the  one-loop
approximation.   Despite significant progress over the last decade
\cite{Fadin:1993ha,Bern:2000ie,Glover:2001ev,Bonciani:2004gi,Bonciani:2004qt,
Bonciani:2005im}, the two-loop corrections have been computed  only in the high
energy limit neglecting the terms suppressed by the  ratio of the electron mass
$m_e$ to the center-of-mass energy $\sqrt{s}$
\cite{Penin:2005kf,Penin:2005eh,Actis:2007gi,Becher:2007cu,Bonciani:2007eh,
Bonciani:2008ep, Actis:2007fs,Kuhn:2008zs}.\footnote{For a review see
Ref.~\cite{Actis:2010gg}} The logarithmically enhanced two-loop electroweak
corrections are available in this approximation as well
\cite{Kuhn:1999nn,Kuhn:2001hz,Feucht:2004rp,Jantzen:2005az,Penin:2011aa}. At the
same time the power-suppressed terms in two loops are still beyond the reach of
existing computational techniques. In general the power-suppressed contributions
are of great interest. At the intermediate energies the power corrections in
many cases are phenomenologically important. Moreover, in contrast to the
leading-power contribution  very little is known about the infrared structure of
the power-suppressed terms. This problem has been studied already in early days
of QED \cite{Gorshkov:1966ht} and currently attracts much attention in various
context
\cite{Kotsky:1997rq,Laenen:2010uz,Banfi:2013eda,Penin:2014msa,Melnikov:2016emg}.
However, a systematic renormalization group analysis of the high-energy behavior
of the on-shell amplitudes beyond  the leading-power approximation is still
elusive for the existing effective field theory methods.

In this paper we  consider  the   ${\cal O}(m_e^2/s)$ two-loop QED corrections
to the differential cross section of the high-energy large-angle Bhabha
scattering. The corrections are evaluated in the double-logarithmic
approximation {\it i.e.} retaining the terms  enhanced by  two powers of the
large logarithm $\ln(s/m_e^2)$ per each power of the coupling constant. These
terms dominate the power-suppressed contribution and in  a wide  energy interval
are numerically comparable to the nonlogarithmic leading-power terms. The
leading power-suppressed double-logarithmic  corrections have been obtained in
Ref.~\cite{Penin:2014msa} to all orders in fine structure constant $\alpha$ for
the electromagnetic form factor of electron. In this paper we elaborate the
approach~\cite{Penin:2014msa} and apply it to the electron-positron scattering
amplitude in two-loop approximation. Our main result  is given by
Eq.~(\ref{eq::result}).

The paper is organized as follows. In the next section we describe the
perturbative expansion of the cross section at high energy. In
Sect.~\ref{sec::3} we discuss the origin and general structure of the
double-logarithmic corrections. In  Sect.~\ref{sec::4} we describe  the
evaluation  of the one and two-loop double-logarithmic power-suppressed
corrections to Bhabha scattering.  Sect.~\ref{sec::summary} is our summary and
conclusion.


\section{Perturbative expansion of the cross section at high energy}
\label{sec::2}
We consider the electron-positron scattering $e^-(p_1)e^+(p_2)\to
e^-(p_3)e^+(p_4)$ at high energy and large angle when all the kinematic
invariants $s_{ij}=(p_i+p_j)^2$ for $i\ne j$ are of the same scale much larger
than  $m_e^2$.\footnote{ All the external momenta are defined  to be incoming
and on-shell so that $p_i^2=m_e^2$ and the Mandelstam variables are $s=s_{12}$,
$t=s_{13}$, and  $u=s_{14}$.} In this limit the cross section can be written as
a series in a small ratio $\rho=m^2_e/s$
\begin{equation}
\sigma={\alpha^2\over s}\sum_{n=0}^\infty\rho^n\sigma_n\,,
\label{eq::rhoseries}
\end{equation}
where  $\sigma_n$ are the functions of $x=-t/s\sim 1$.\footnote{The variable $x$
is related to the scattering angle $\theta$ in the center of mass frame,
$x=(1-4\rho)(1-\cos\theta)/2$.} These functions  in turn  can be computed as
series in $\alpha$.  Up to ${\cal O}(\alpha)$ the result for the cross section
is known in a closed analytical  form (see {\it  e.g.} \cite{Bonciani:2004gi})
and the coefficients  in Eq.~(\ref{eq::rhoseries}) can be found for any $n$.
The second order result is available only for the leading-power contribution
$\sigma_0$.  The series~(\ref{eq::rhoseries}) is asymptotic  and after the
expansion in $\alpha$ its  coefficients in general have logarithmic dependence
on  $\rho$.  In the high-energy limit the double-logarithmic contributions
enhanced by  two powers of the large logarithm $\ln\rho$ per each power of the
coupling constant dominate the expansion of $\sigma_n$  in $\alpha$. In the
double-logarithmic approximation perturbative expansion for these coefficients
can be written as series in  $\tau = {\alpha\over 4\pi}\ln^2\rho$
\begin{equation}
\sigma_n=\exp{\left[-{2\alpha\over \pi}B(\rho)
\ln\left({\lambda^2/m_e^2}\right)\right]}
\sum_{m=0}^\infty \tau^m\sigma^{(m)}_n\,.
\label{eq::tauseries}
\end{equation}
In Eq.~(\ref{eq::tauseries}) the exponential prefactor  with
$B(\rho)=\ln\rho+{\cal O}(1)$ accounts for the universal singular dependence of
the amplitude on the auxiliary photon mass $\lambda$ introduced to regulate the
infrared divergences \cite{Yennie:1961ad}.  For the  leading-power term the
double-logarithmic corrections are know to factorize and exponentiate
\cite{Sudakov:1954sw,Jackiw:1968zz,Gorshkov:1973if,Frenkel:1976bj,Mueller:1979ih,
Collins:1980ih,Sen:1981sd,Sen:1982bt,Sterman:1986aj}. In this approximation  the
all-order dependence of the differential cross section on $\tau$  is given by the
expression
\begin{equation}
{{\rm d}\sigma_0\over{\rm d}\Omega}=
e^{-4\tau}
{{\rm d}\sigma^{(0)}_0\over{\rm d}\Omega}\,,
\label{lpdl}
\end{equation}
where the Born term reads
\begin{equation}
{{\rm d}\sigma^{(0)}_0\over{\rm d}\Omega}
=\left({1-x+x^2\over x}\right)^2\,.
\label{eq::lpborn}
\end{equation}
The goal of this paper is to compute the coefficient $\sigma^{(2)}_1$.


\section{General structure of double-logarithmic corrections}
\label{sec::3}
The double-logarithmic terms are in general associated with the soft and
collinear  divergences of the amplitudes due to radiation  of the soft virtual
particles  by  highly energetic on-shell charged particles. At the same time the
structure of the double-logarithmic corrections  crucially depends on their
origin. Below we consider  two types of the double-logarithmic corrections,
which play the central role in our analysis.

\subsection{Sudakov logarithms}
Sudakov double-logarithmic corrections are induced by  the soft photon exchange.
In the leading order of the high energy/small mass expansion the Sudakov double
logarithms exponentiate and result in a strong universal suppression of any
electron scattering amplitude with a fixed number of emitted photons in the
limit when all the kinematic invariants of the process are large,
Eq.~(\ref{lpdl}). A crucial  observation of  Ref.~\cite{Penin:2014msa} is that
``Sudakov'' photons do not generate ${\cal O}(\rho)$ double-logarithmic
corrections to the scattering  amplitudes. Below we present a detailed
derivation of this result.

Let us outline our approach to the analysis of the power-suppressed
double-logarithmic contributions. We use the expansion by regions method
\cite{Smirnov:1997gx,Smirnov:2002pj} to get a systematic expansion of the
Feynman integrals in $\rho$. Within this method every  Feynman integral is given
by the sum over contributions of different virtual momentum regions. Each
contribution is represented by a homogeneous Feynman integral, which in general
is divergent even if the original integral before the expansion is finite. These
spurious divergences result from the process of  scale separation and have to be
dimensionally regulated. The singular terms cancel out in the sum of all regions
but can be used to find the logarithmic terms.  The double-logarithmic
contributions are determined by the leading singular behavior of the integrals
and can be found by the method developed in Ref.~\cite{Sudakov:1954sw} (see also
\cite{Gorshkov:1966ht,Gorshkov:1973if}). Though the method is blind to the power
corrections, it can be applied in this case since the expansion by regions
provides the integrals, which are {\it homogeneous} in the expansion parameter.
Let us consider first an exchange of a virtual photon with the momentum $l$
between on-shell fermion lines with the momenta $p_i$ and $p_j$. The Sudakov
double logarithmic contribution originates from the region where the photon
momentum is small. Thus we can neglect it in the numerator of the fermion
propagators since the integral with the additional power of the photon momentum
is not sufficiently singular to develop the double-logarithmic behavior. Then by
using the equations of motion $(\slashed{p}_i-m_e)\psi(p_i)=0$ the soft photon
contribution can be reduced to the integral
\begin{equation}
I=\int {{\rm d}^dl} {(p_ip_j)\over l^2((p_i-l)^2-m_e^2)( (p_j+l)^2-m_e^2)}\,.
\label{eq::triangle}
\end{equation}
In the above equation we neglected the photon mass and use the dimensional
regularization with $d=4-2\varepsilon$. The soft divergence in this case appear
as a pole in $\varepsilon$. This modifies the form of the exponent in
Eq.~(\ref{eq::tauseries}) but does not affect the structure of the expansion in
$\rho$. The integral gets contributions from the hard and two (symmetric)
collinear regions $I=I_h+I_{c-i}+I_{c-j}$. Since the singularities of the hard
and collinear regions are not independent, it is sufficient to consider only the
contribution of a single region, {\it e.g.} the $i$-collinear one $I_{c-i}$. We
set the parameter of dimensional regularization to be $\mu^2\sim s_{ij}$, so
that the expansion of the hard region  contribution with $l\sim \sqrt{s_{ij}}$
in  $\varepsilon$  does not produce large logarithms. For the  large-angle
scattering we can choose the light-cone coordinates where $p_1\approx {p_i}_-$
and $p_{j}\approx{p_j}_+$. Then the  $i$-collinear region is defined by the
following scaling of the virtual momentum components  $l_+\sim
m_e^2/\sqrt{s_{ij}},~l_-\sim \sqrt{s_{ij}},~l_\perp\sim m_e$, so that $\l^2\sim
m_e^2$. It is convenient to introduce the light-like vectors $\tilde
p_{i},~\tilde p_{j}$ such that $p_i=\tilde p_i+{m_e^2\over\tilde s_{ij}}\tilde
p_j$ and $p_j=\tilde p_j+{m_e^2\over \tilde s_{ij}}\tilde p_i$, where   $(\tilde
p_i\tilde p_j)=\tilde s_{ij}$. In the $i$-collinear region the electron
propagator is substituted by the series
\begin{equation}
{1\over (p_j+l)^2-m_e^2}=\sum_{n=0}^\infty(-1)^n
{(2(m_e^2/\tilde s_{ij})(l\tilde{p}_i)+l^2)^n\over (l\tilde{p}_j)^{(n+1)}}\,,
\label{eq::expprop}
\end{equation}
which results in  a series
\begin{equation}
I_{c-i}=\int {{\rm d}^dl} {(p_ip_j)\over l^2(-2(p_il)+l^2) (l\tilde{p}_j)}
\left[1-{2(m_e^2/\tilde s_{ij})(l\tilde{p}_i)+l^2\over (l\tilde{p}_j)}
+{\cal O}(m_e^4/\tilde s^2_{ij})\right]\,.
\label{eq::icseries}
\end{equation}
Let us consider  the second  term in Eq.~(\ref{eq::icseries}).  In the limit
when the virtual momentum is soft and collinear to $p_i$ either $l^2$ or $p_il$
factor in the denominator is cancelled and the integrand is therefore not
singular enough to develop the  double-logarithmic contribution.  At the same
time by integrating the first term one gets
\begin{equation}
(i\pi^2){2(p_ip_j)\over \tilde s_{ij}}
\left[-{1\over \varepsilon}\ln\left({m_e^2\over\tilde s_{ij}}\right)
+{1\over 2}\ln^2\left({m_e^2\over\tilde s_{ij}}\right)
\right]\,,
\label{eq::iclo}
\end{equation}
where only the  double-logarithmic contribution is retained and the pole
corresponds to the soft divergence not regulated by the electron mass. Since
${2(p_ip_j)/ \tilde s_{ij}}=1+{\cal O}(m_e^4/\tilde s^2_{ij})$,
Eq.~(\ref{eq::iclo}) can be written as follows
\begin{equation}
(i\pi^2)
\left[-{1\over \varepsilon}\ln\rho+{1\over 2}\ln^2\rho+{\cal O}(\rho^2)
\right]\,,
\label{eq::iclorho}
\end{equation}
{\it i.e.} the first term of the expansion~(\ref{eq::icseries}) does not
generate ${\cal O}(\rho)$ double-logarithmic corrections as well. The above
analysis can be generalized to an arbitrary number of Sudakov photons.  After
neglecting all the Sudakov photon momenta in the numerators the Lorentz/spinor
reduction becomes straightforward.   By using the equations of motion and the
on-shell conditions one gets the factor $(p_ip_j)$ per each photon connecting
the lines with the momenta $p_i$ and $p_j$ for any $i$ and $j$. At the same time
 the structure of the expansion by regions becomes  more involved.  For the
multiloop diagrams it also includes  ultra-collinear regions, which are obtained
by multiplying   the collinear scaling rules with a power of $(m_e^2/\tilde
s_{ij})$.  All these  regions should be taken into account to find the total
double-logarithmic contribution.  As an example let us consider all the virtual
momenta  $l_k$ to be  $i$-collinear. It represents  the most complicated case
since the integrations over  different $l_k$ do not factorize. After the
expansion one gets eikonal propagators of the form~(\ref{eq::expprop}), which
depend on a sum of  several virtual momenta $l_k$ with identical scaling. Since
the expansion by regions generates homogeneous integrals, the leading term of
the expansion is proportional to a product  of $(p_ip_j)/\tilde s_{ij}$ factors
for different $i$ and $j$ and therefore does not produce any  ${\cal O}(\rho)$
terms. Then for the analysis of the next-to-leading term  we use the
method~\cite{Sudakov:1954sw} to extract the double-logarithmic asymptotic
behavior of a given integral. According
to~\cite{Sudakov:1954sw,Gorshkov:1966ht,Gorshkov:1973if} the double-logarithmic
contribution originates from the region of strongly ordered virtual momenta
determined by a set of conditions $(l_{k_1}p_m)\ll(l_{k_2}p_m)\ll \ldots \ll
(l_{k_n}p_m)$ for any $m$ and some permutations of the indices $k_i$.  Thus in
the double-logarithmic region one can neglect all the virtual momenta but one in
each eikonal propagator and the problem effectively reduces to the one-loop case
considered above, where the other virtual momenta only play a role of an
infrared or ultraviolet cutoff for the double-logarithmic integration. Due to  a
natural ordering of the momenta with different collinearity the analysis of the
double-logarithmic contribution of the corresponding mixed regions does not
differ from the case considered above.

Thus we have found that Sudakov photons do not produce  double-logarithms in the
first order in $\rho$. We have  checked the absence of the   ${\cal O}(\rho)$
double-logarithmic contribution by explicit evaluation of the collinear region
contributions to the two-loop scalar integrals, which appear in the analysis of
the Bhabha scattering. This observation agrees with the analysis
\cite{Korchemsky:1987wg} of the cusp anomalous dimension, which determines the
double-logarithmic corrections to the light-like Wilson line with a cusp. For
the large cusp angle corresponding to the limit $\rho\to 0$  from the result of
Ref.~\cite{Korchemsky:1987wg} one gets
\begin{equation}
\Gamma_{cusp}=-{\alpha\over\pi}\ln\rho\left(1+{\cal O}(\rho^2)\right),
\end{equation}
with vanishing first-order term in $\rho$. Our result,  however, is more general
since it also implies the absence of ``kinematic'' ${\cal O}(\rho)$ corrections,
which multiply the leading-order cusp anomalous dimension when the scattering
amplitude  is related to the Wilson line.

Note that the double-logarithmic    ${\cal O}(\rho)$  corrections do vanish only
for the {\it amplitudes}. When the amplitudes are squared one gets  ${\cal
O}(\rho)$ terms, which multiply the Sudakov exponential factor and produce the
${\cal O}(\rho)$  double-logarithmic corrections of the form
\begin{equation}
e^{-4\tau}
{{\rm d}\sigma^{(0)}_1\over{\rm d}\Omega}\,.
\label{eq::nlpsud}
\end{equation}

\subsection{Non-Sudakov logarithms}
The ${\cal O}(\rho)$ double-logarithmic contributions to the amplitutes
originate from a completely different virtual momentum configuration. Let us
consider an electron propagator $S(p_i-l)$, where $l$ is the momentum of a
virtual photon with the propagator $D_{\mu\nu}(l)$. In the soft-photon limit
$l\to 0$ the electron propagator becomes eikonal
\begin{equation}
S(p_i-l)\approx-{\slashed{p}_i+m_e\over  2p_il}
\label{eq::sl}
\end{equation}
and develops  a collinear singularity when $l$ is parallel to $p_i$.
Alternatively, we may consider the soft-electron limit $l'\to 0$, where
$l'=p_i-l$. Then the electron propagator becomes scalar
\begin{equation}
S(l')\approx{m_e\over l'^2-m_e^2}
\label{eq::slprime}
\end{equation}
while the photon propagator becomes eikonal
\begin{equation}
D_{\mu\nu}(l')\approx{g_{\mu\nu}\over 2p_il'-m_e^2+\lambda^2}\,.
\label{eq::dlprime}
\end{equation}
Thus the roles of the electron and photon propagators are exchanged. Due to the
explicit factor $m_e$ in  the scalar electron propagator this region can only
generate the mass-suppressed double-logarithmic contribution.  The
existence of non-Sudakov double-logarithmic contributions due to soft electron
exchange has actually been known for a long time \cite{Gorshkov:1966ht}. They
are typical  for the amplitudes that are   mass suppressed  at high energy. In
contrast to the  Sudakov case such logarithms do not factorize and exponentiate.
As a result very little is known about the all-order structure of the
power-suppressed  non-Sudakov logarithms.  Only a few examples of the
non-Sudakov resummation  are  known so far
\cite{Gorshkov:1966ht,Kotsky:1997rq,Penin:2014msa,Melnikov:2016emg}. At the same
time  due to explicit power suppression factor  the soft-electron
double-logarithmic contribution in a given order of perturbation theory can be
determined within the original method of Ref.~\cite{Sudakov:1954sw}.

For the calculation we in general  follow the procedure formulated in
\cite{Penin:2014msa}  for the analysis of the form factor (see also
Ref.~\cite{Melnikov:2016emg}). The structure of the two-loop non-Sudakov
corrections to the electron-positron scattering amplitude has an important
difference though. For the one-loop vertex corrections the virtual momentum
configuration discussed above does not produce a  double-logarithmic
contribution because the  momentum shift distorts  the eikonal structure of the
second electron propagator and removes the soft singularity at small $l'$
necessary to get  the second power of the large logarithm. As a consequence the
${\cal O}(\rho)$ double-logarithmic corrections to the electron form factor
appear first in  two loops in a diagram  with soft electron pair exchange
\cite{Penin:2014msa}. At the same time in the one-loop box diagrams after the
momentum shift both photon propagators become eikonal and provide the necessary
infrared structure. Thus the ${\cal O}(\rho)$ double-logarithmic corrections to
the scattering amplitude appear already in one loop due to a single soft
electron exchange. Therefore for the calculation of the two-loop ${\cal
O}(\rho)$ corrections to the Bhabha scattering one has to take into account  the
diagrams with both  soft electron and soft photon in addition to the soft
electron pair contribution. We discuss the details of the calculation in the
next section.


\section{Double-logarithmic  ${\cal O}(\rho)$  corrections to Bhabha scattering}
\label{sec::4}
In the analysis of the Feynman diagrams we always choose the momentum routing in
such a way that a soft electron or photon line carries a single virtual momentum
only. For the determination of the double-logarithmic contribution we can use
the  effective Feynman rules, which retain the leading infrared behavior of the
full theory. For a soft electron line we make the following approximation
\begin{equation}
\frac{\hat l + m_e }{l^2 - m_e^2} \to \frac{m_e}{l^2 - m_e^2}\,,
\end{equation}
so that it effectively becomes scalar. For an electron carrying a single
external momentum we use the eikonal approximation
\begin{equation}
\frac{ \hat p_i+ \hat l +m_e}{(p_i+l)^2 - m_e^2} \to
\frac{\hat p_i +m_e}{2 (p_il)}\,.
\end{equation}
An electron line with  two different external momenta corresponds to  a far
off-shell or ``hard'' electron propagator
\begin{equation}
\frac{\hat p_i+\hat p_j+ \hat l +m_e}{(p_i+p_j+l)^2 - m_e^2} \to
\frac{\hat p_i+\hat p_j +m_e}{s_{ij}}\,,
\end{equation}
which effectively  reduces to a local interaction vertex. Similar approximation
is used for the eikonal and hard photons
\begin{equation}
\frac{g_{\mu\nu}}{(p_i+l)^2-\lambda^2}\to\frac{g_{\mu\nu}}{2 (p_il)+m_e^2-\lambda^2},
\qquad
\frac{g_{\mu\nu}}{(p_i+p_j+l)^2 - \lambda^2} \to \frac{g_{\mu\nu}}{s_{ij}}.
\end{equation}
In principle the Feynman rules can be further simplified by using the light-cone
coordinates. In this case the soft  photon and eikonal electon propagators
have only the light-cone components, and so on. In two loops, however, the
standard tools for Lorentz and spinor algebra turn out to be more convenient.

The next big simplification is related to the treatment of the  dependence of 
the corrections on the photon mass. In the diagrams without soft photon lines 
both soft and collinear divergences of the virtual momentum integration are 
regulated by the electron mass. These diagrams are not sensitive to the photon 
mass and in the double-logarithmic approximation can be computed either with 
$\lambda=0$ or $\lambda=m_e$ with the identical result. The diagrams with both 
soft photon and soft electron exchanges do depend  on $\lambda$. This 
dependence, however, can be determined from the general properties of soft 
photon contribution. Indeed, the virtual momentum space of the soft photons with 
$|{\bf l}|\ll m_e$ is known to factorize \cite{Yennie:1961ad}. For high-energy 
scattering the integration over such momenta results in the exponent of the 
one-loop contribution in Eq.~(\ref{eq::tauseries}). If we perform the 
calculation with $\lambda \sim m_e$,  this part of the virtual momentum space is 
eliminated so that the exponent in Eq.~(\ref{eq::tauseries}) reduces to a 
nonlogarithmic factor and we directly obtain the coefficients $\sigma_n^{(m)}$. 
In this way we reduce the number of different scales in the problem, which 
significantly simplifies the  analysis. It is important to note that the above 
factorization works only for the sum of a given class of the diagrams. The 
remaining  infrared finite  diagrams may have different double-logarithmic 
behavior  for  $\lambda=0$ and  $\lambda=m_e$ and should be computed with 
massless photon.

\subsection{One-loop  contributions}
According to the discussion of Sect.~\ref{sec::3} the one-loop leading-power
corrections have  two distinct sources. The soft photon part is determined by
the product of the standard Sudakov double-logarithmic corrections to the
scattering amplitudes and the ${\cal O}(\rho)$ Born cross section. It  is given
by  the first term of the expansion of Eq.~(\ref{eq::nlpsud}) in $\tau$. The
non-Sudakov contribution is generated by the box diagrams with one soft and one
hard electron line and two eikonal photon propagators. We compute it by using
the effective Feynman rules introduced in the previous section. The total result
for the one-loop double-logarithmic power-suppressed contribution is
\begin{equation}
{{\rm d}\sigma^{(1)}_1\over{\rm d}\Omega}
=-4{{\rm d}\sigma^{(0)}_1\over{\rm d}\Omega}+
{6-20x+24x^2-20x^3+6x^4\over (1-x)x^2}
\,,
\label{eq::oneloop}
\end{equation}
where the first and the second terms correspond to the soft photon and soft
electron contributions, respectively.  It agrees with the known analytic
one-loop result~\cite{Bonciani:2005im} expanded to  ${\cal O}(\rho)$.

\begin{figure}[t]
\begin{center}
\begin{tabular}{ccc}
\includegraphics[width=3cm]{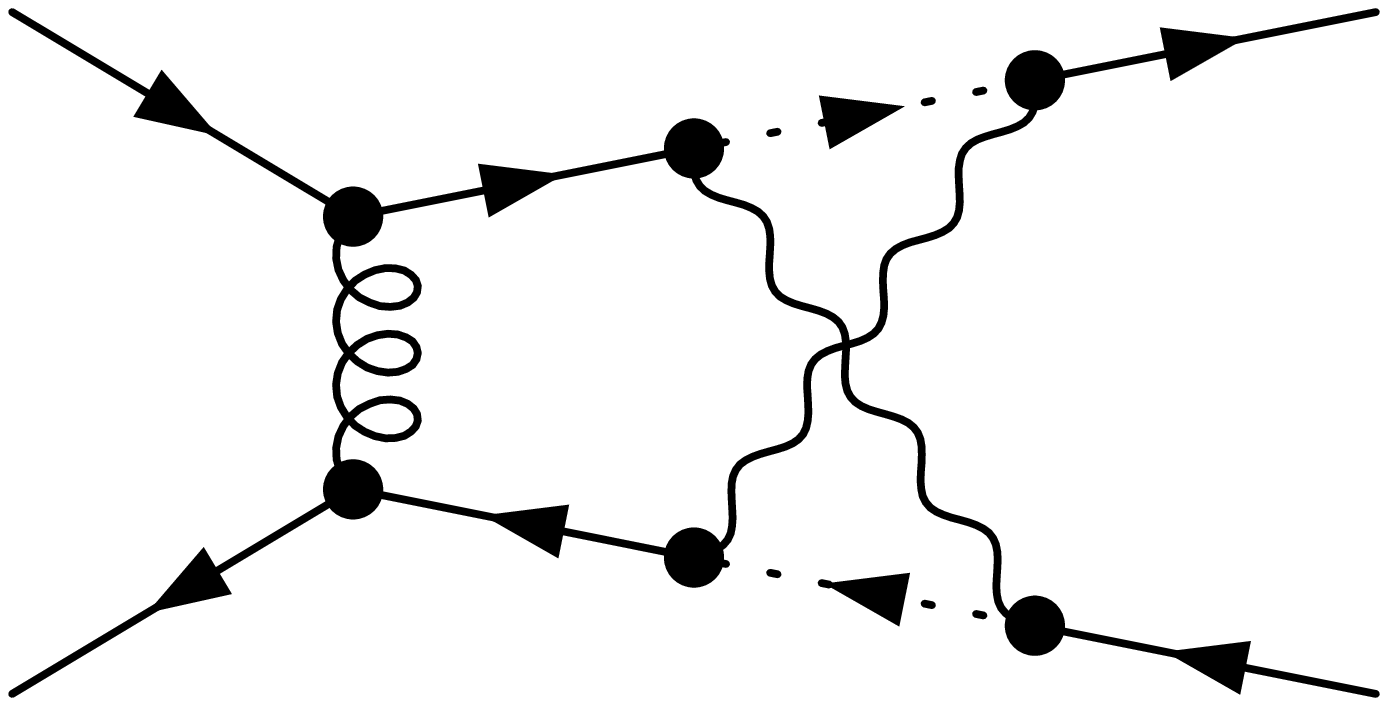}&
\includegraphics[width=3cm]{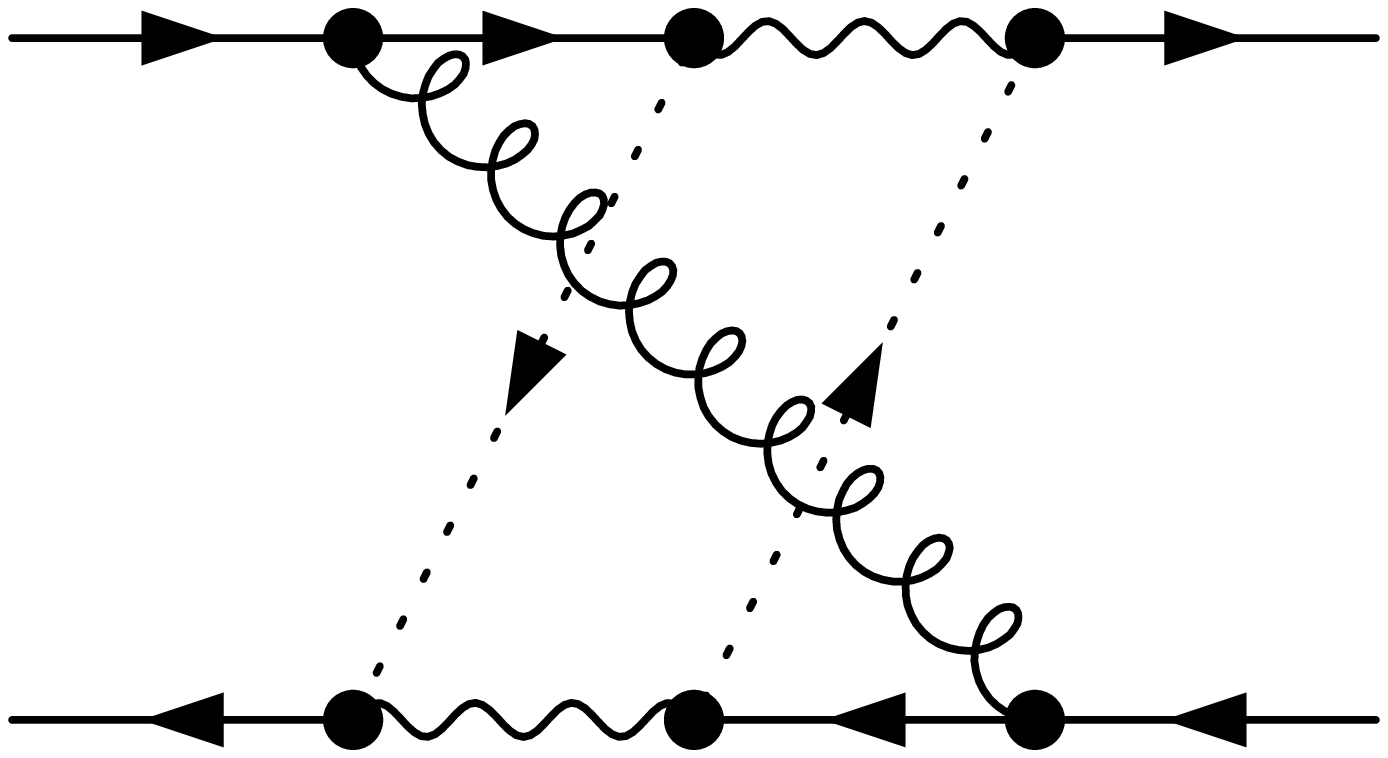}&
\includegraphics[width=3cm]{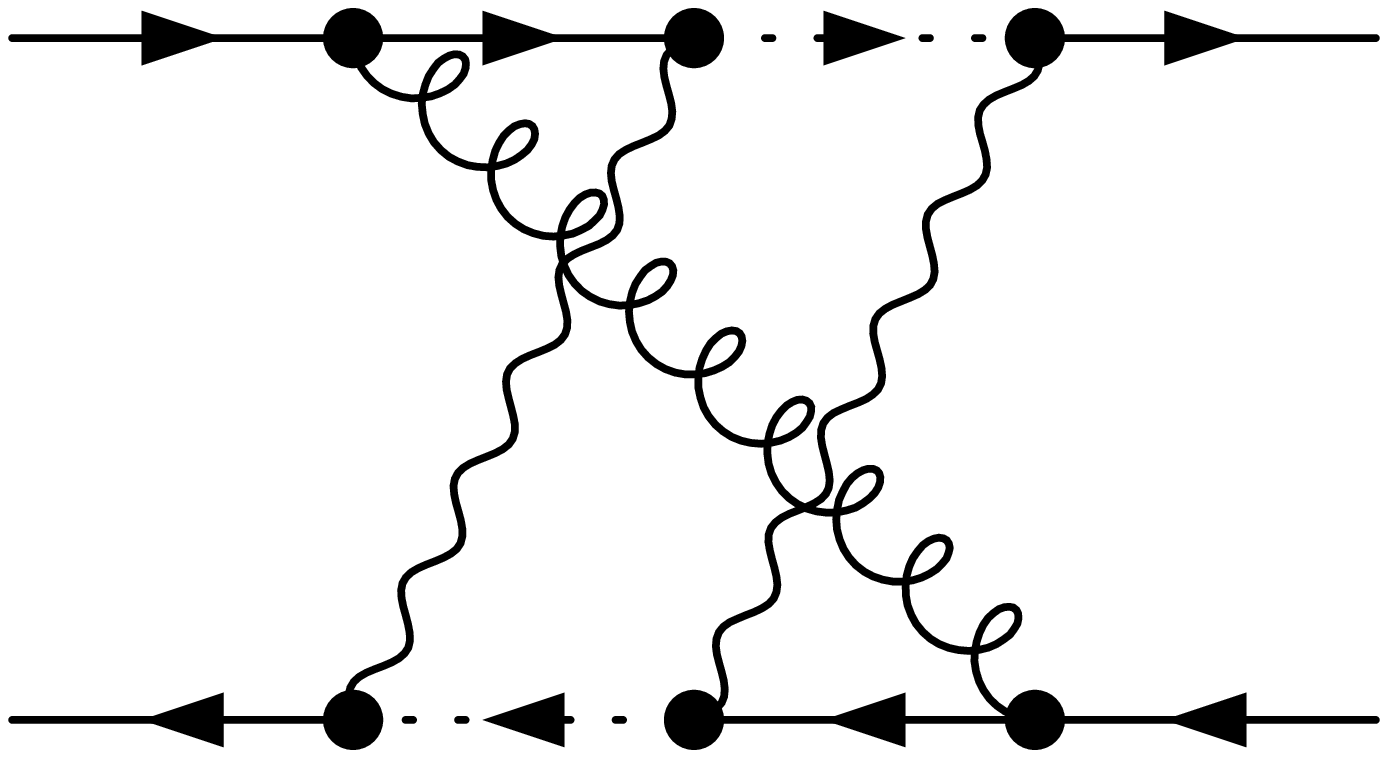}\\[-8mm]
(a)&(b)&(c)\\
&&\\
\includegraphics[width=3cm]{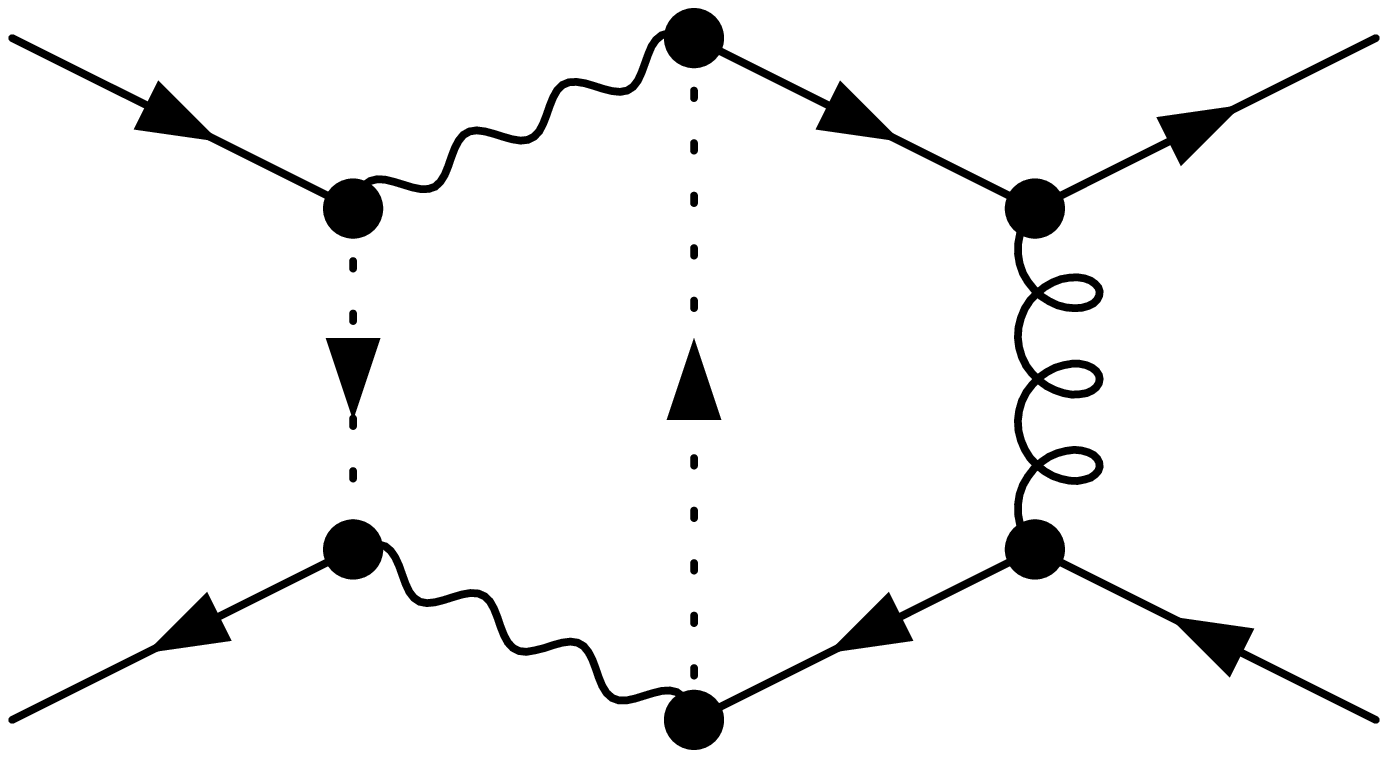}&
\includegraphics[width=3cm]{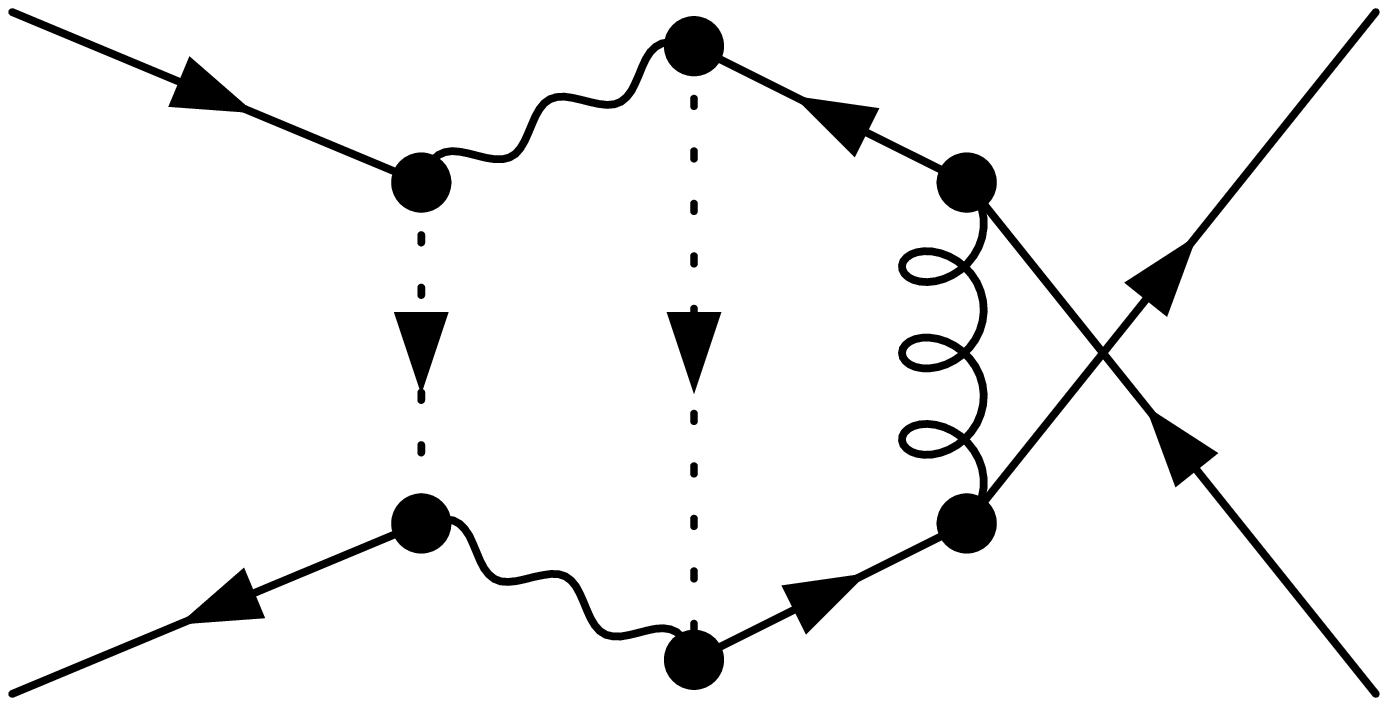}&
\includegraphics[width=3cm]{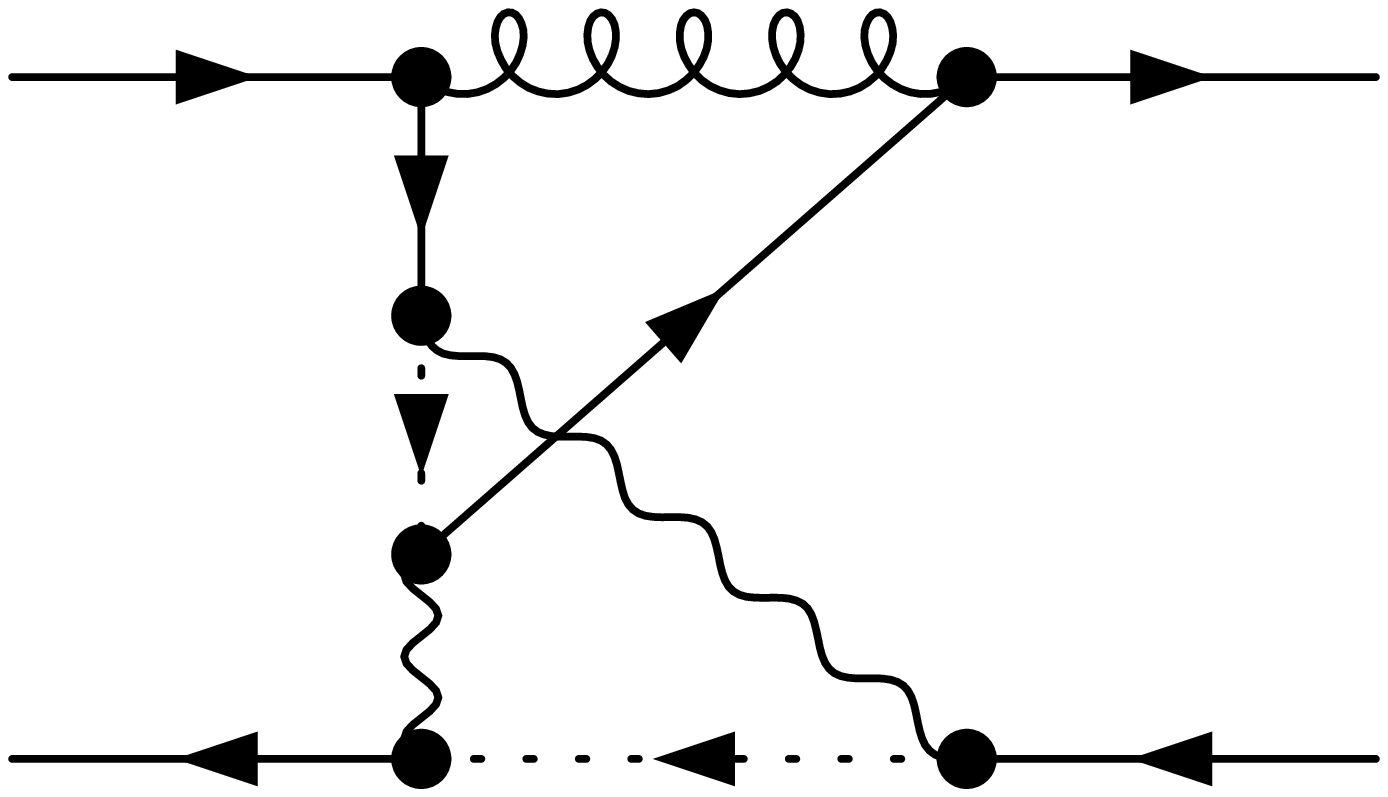}\\[-8mm]
(d)&(e)&(f)
\end{tabular}
\end{center}
\caption{\label{fig::1}  Two-loop  one-particle irreducible diagrams  with soft
electron pair exchange.  Dashed arrows correspond to the scalar soft quarks. The
loopy (wavy) lines correspond to  the hard (eikonal) photons. Symmetric diagrams
are not shown. }
\end{figure}

\subsection{Two-loop  contributions}
In two loops the double-logarithmic power-suppressed contribution can be
decomposed as follows
\begin{equation}
{{\rm d}\sigma^{(2)}_1\over{\rm d}\Omega}
={{\rm d}\sigma^{(2)}_1\over{\rm d}\Omega}\Bigg|_{1l\times 1l}
+{{\rm d}\sigma^{(2)}_1\over{\rm d}\Omega}\Bigg|_{1PR}+
{{\rm d}\sigma^{(2)}_1\over{\rm d}\Omega}\Bigg|_{1PI}\,,
\label{twoloop}
\end{equation}
where three terms correspond to  the one-loop by one-loop amplitude
interference, the two-loop one-particle reducible and  one-particle irreducible
corrections to the amplitude, respectively. The calculation of the interference
term is straightforward and gives
\begin{equation}
{{\rm d}\sigma^{(2)}_1\over{\rm d}\Omega}\Bigg|_{1l\times 1l}
=-4{{\rm d}\sigma^{(0)}_1\over{\rm d}\Omega}-2
{{\rm d}\sigma^{(1)}_1\over{\rm d}\Omega}
\,.
\label{eq::1lsquare}
\end{equation}
The  two-loop one-particle reducible contribution is determined by the
corrections to the electron form factor. Its soft photon part  is given by the
interference of the two-loop Sudakov form factor and square of the one-loop
Sudakov form factor  with  the ${\cal O}(\rho)$  part of the Born cross
section. The non-Sudakov corrections are generated by the two-loop soft electron
pair exchange and can be found in  \cite{Penin:2014msa} (see also
\cite{Mastrolia:2003yz,Bernreuther:2004ih} for the full theory calculation).
The total reducible contribution reads
\begin{equation}
{{\rm d}\sigma^{(2)}_1\over{\rm d}\Omega}\Bigg|_{1PR}
=4{{\rm d}\sigma^{(0)}_1\over{\rm d}\Omega}-
{4-6x+8x^2-8x^3+6x^4-4x^5\over 3x^3}
\,,
\label{eq::1pr}
\end{equation}
where the  first and the second terms correspond to the soft photon and soft
electron pair contributions, respectively.

\begin{figure}[t]
\begin{center}
\begin{tabular}{cccc}
\includegraphics[width=3cm]{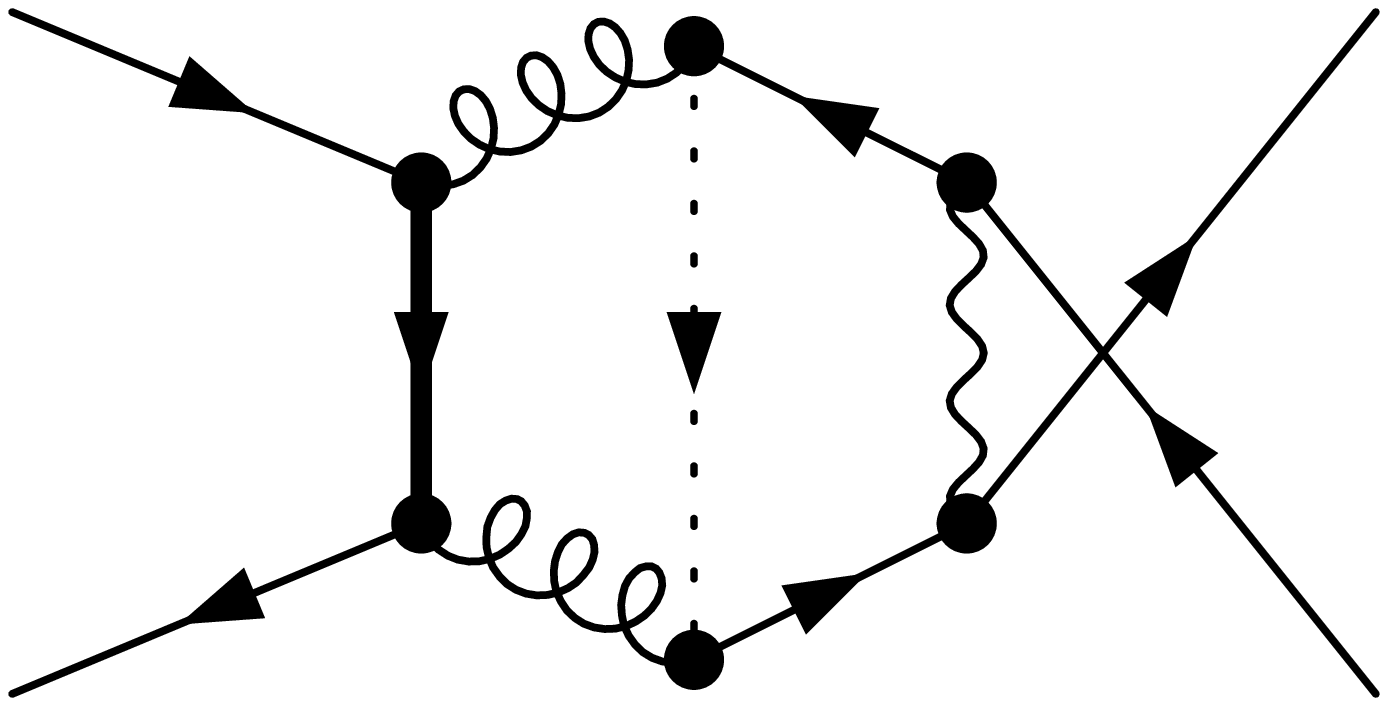}&
\includegraphics[width=3cm]{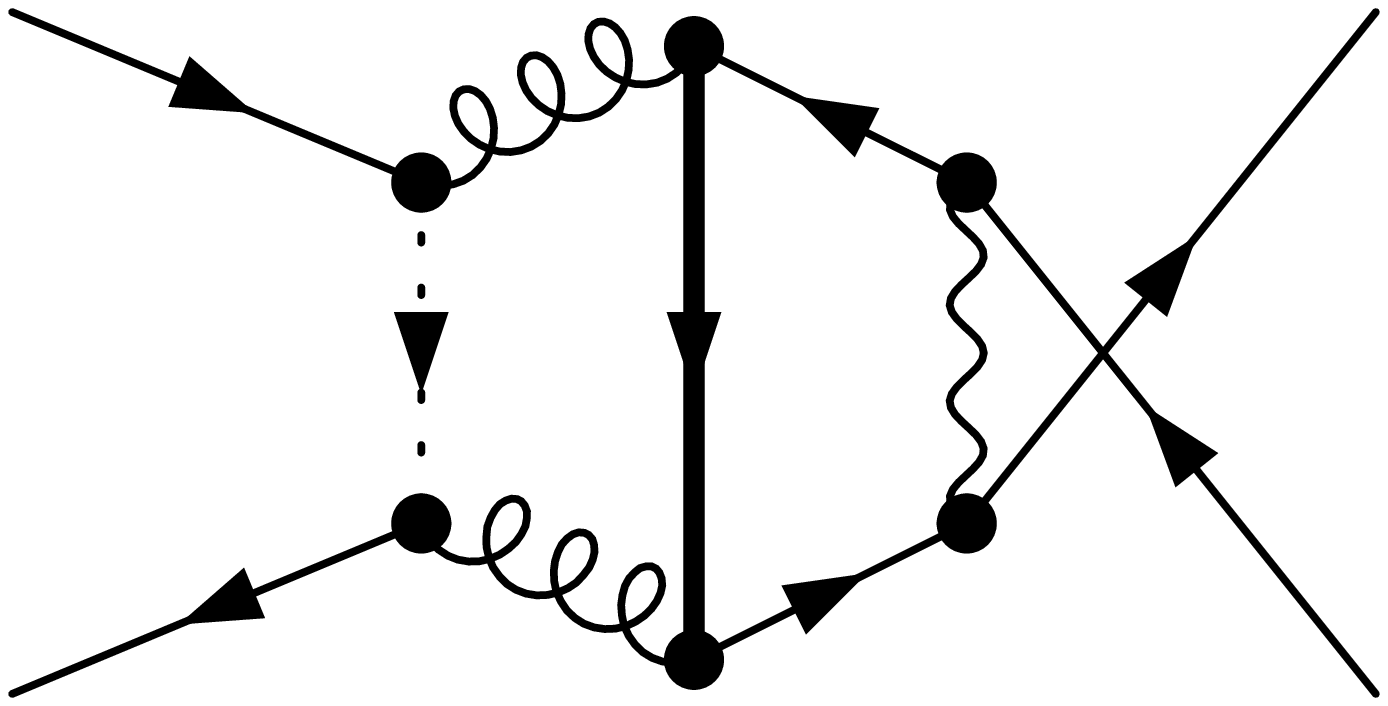}&
\includegraphics[width=3cm]{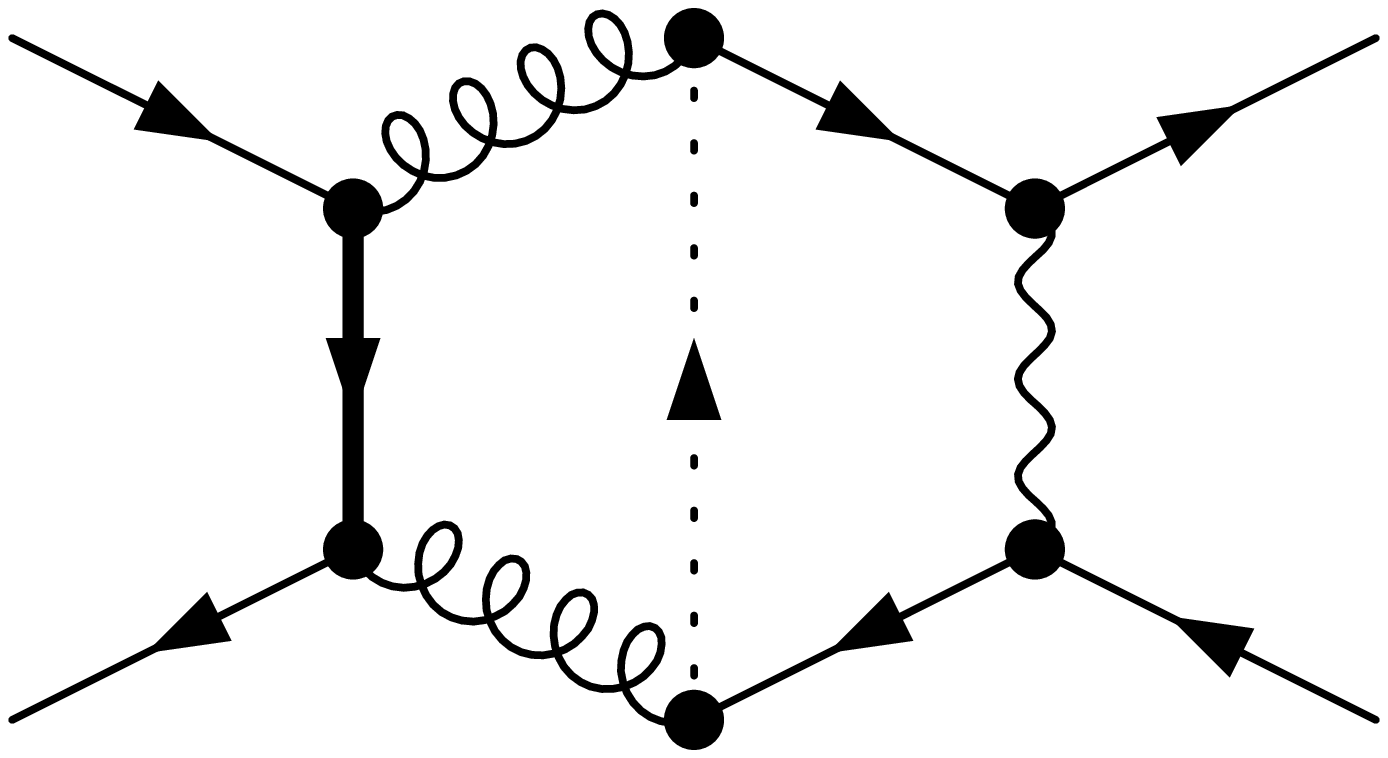}&
\includegraphics[width=3cm]{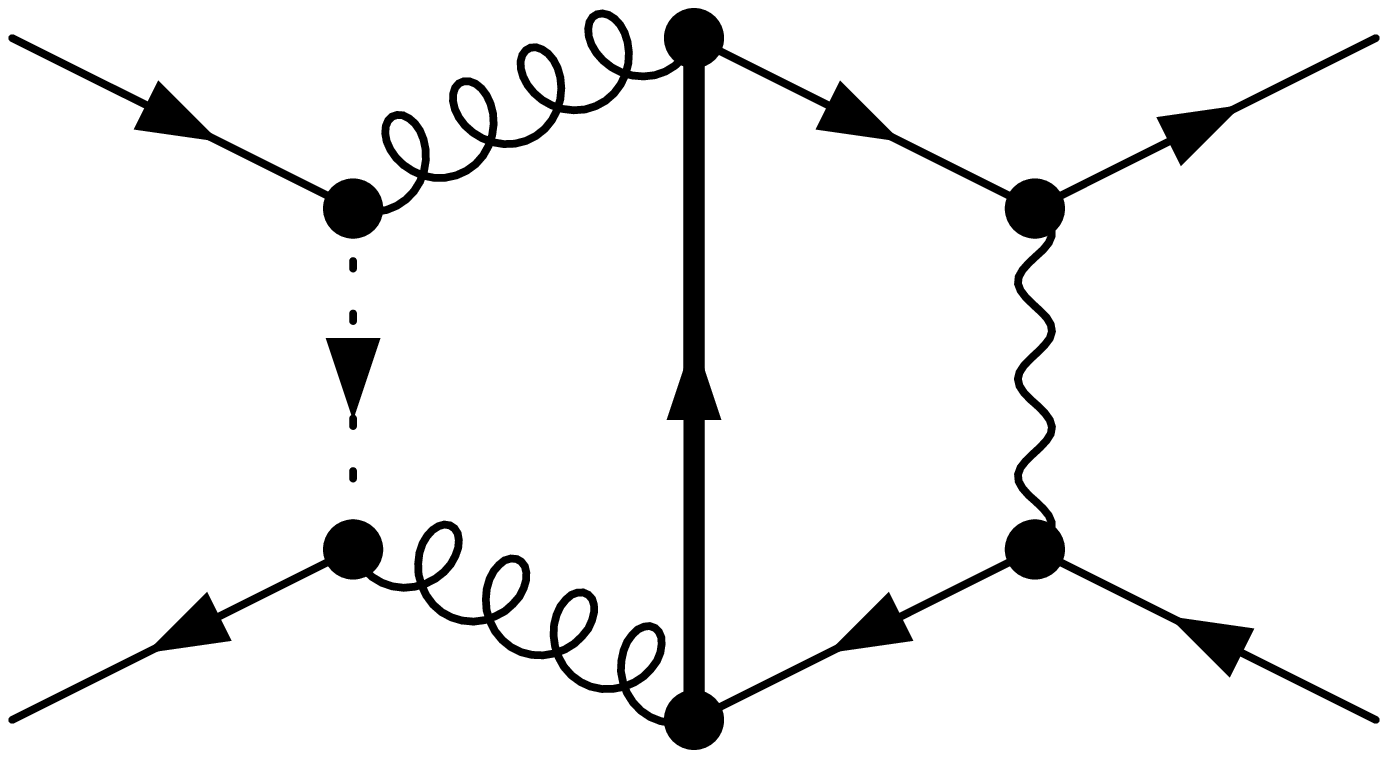}
\\[-8mm]
(a)&(b)&(c)&(d)\\
\includegraphics[width=3cm]{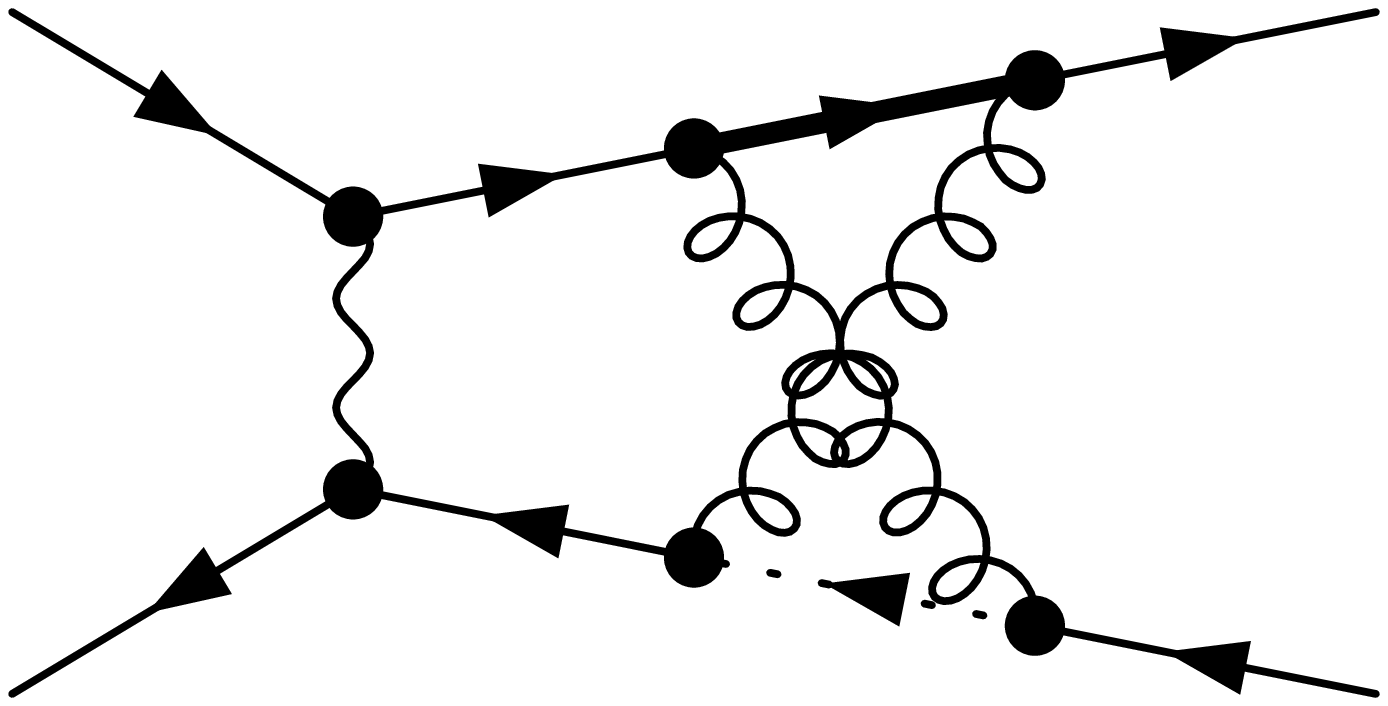}&
\includegraphics[width=3cm]{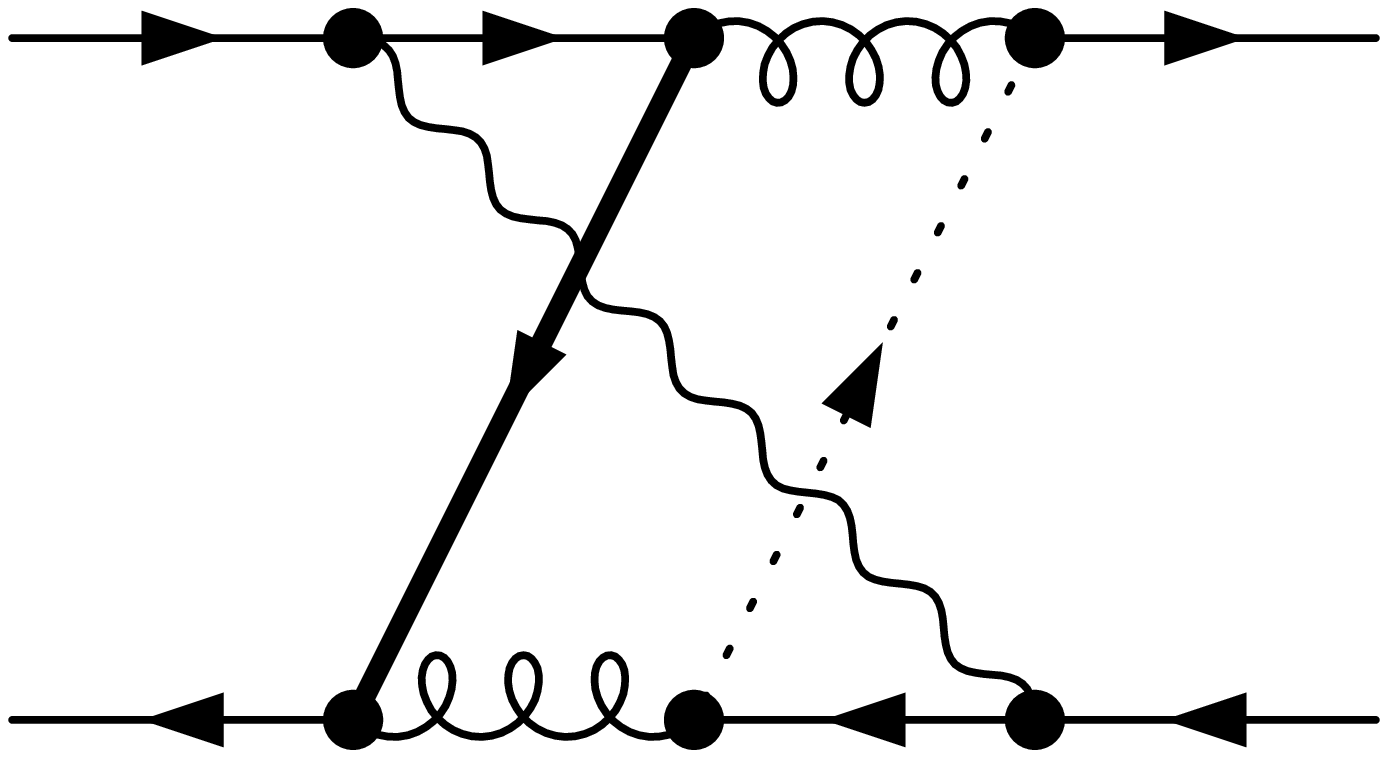}&
\includegraphics[width=3cm]{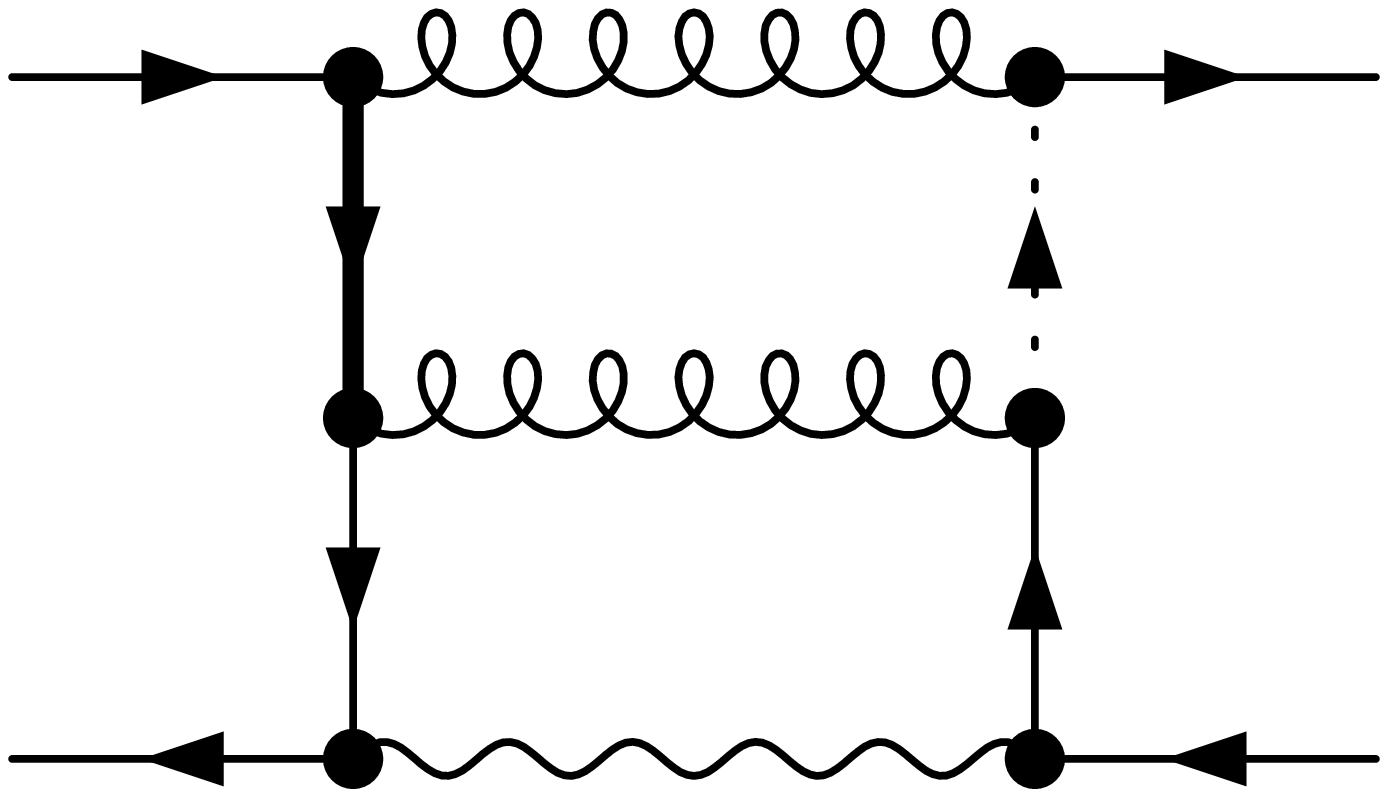}&
\includegraphics[width=3cm]{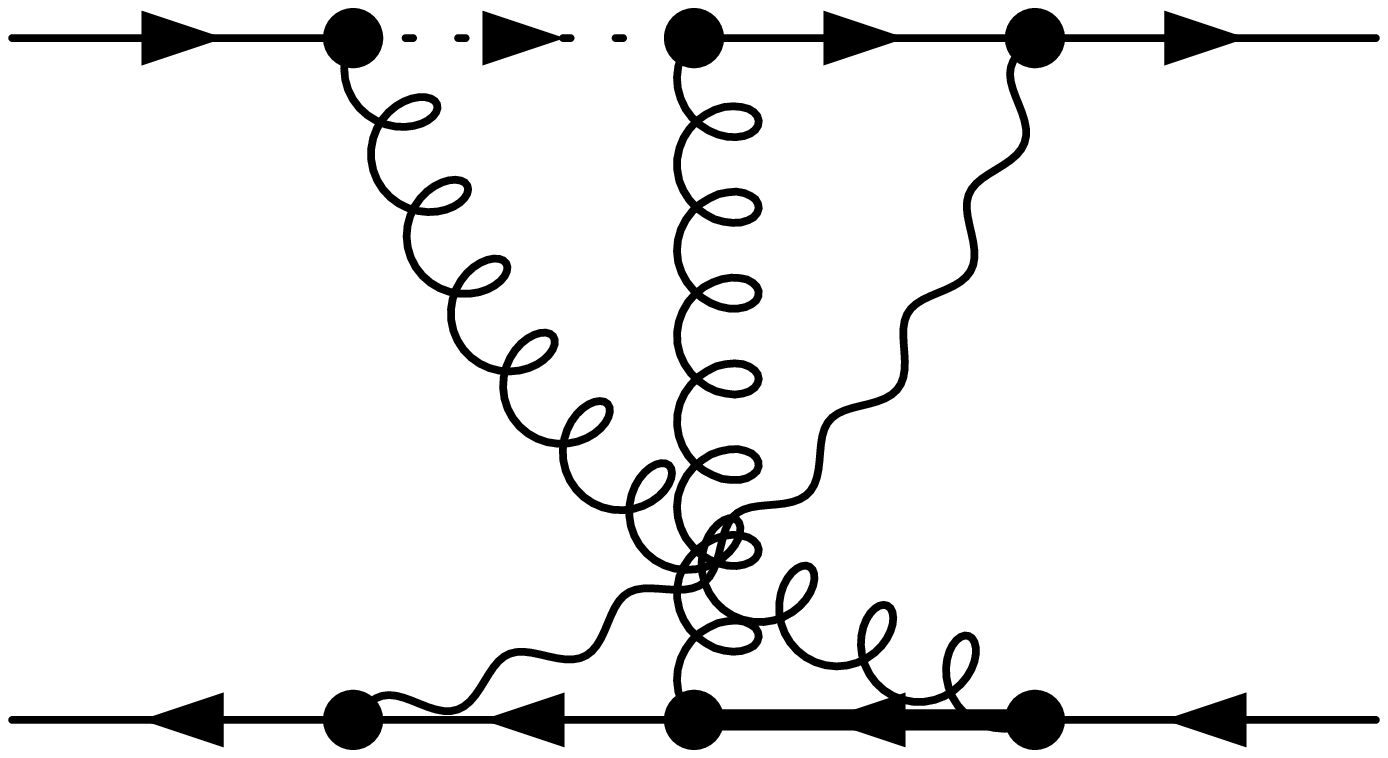}
\\[-8mm]
(e)&(f)&(g)&(h)\\
\includegraphics[width=3cm]{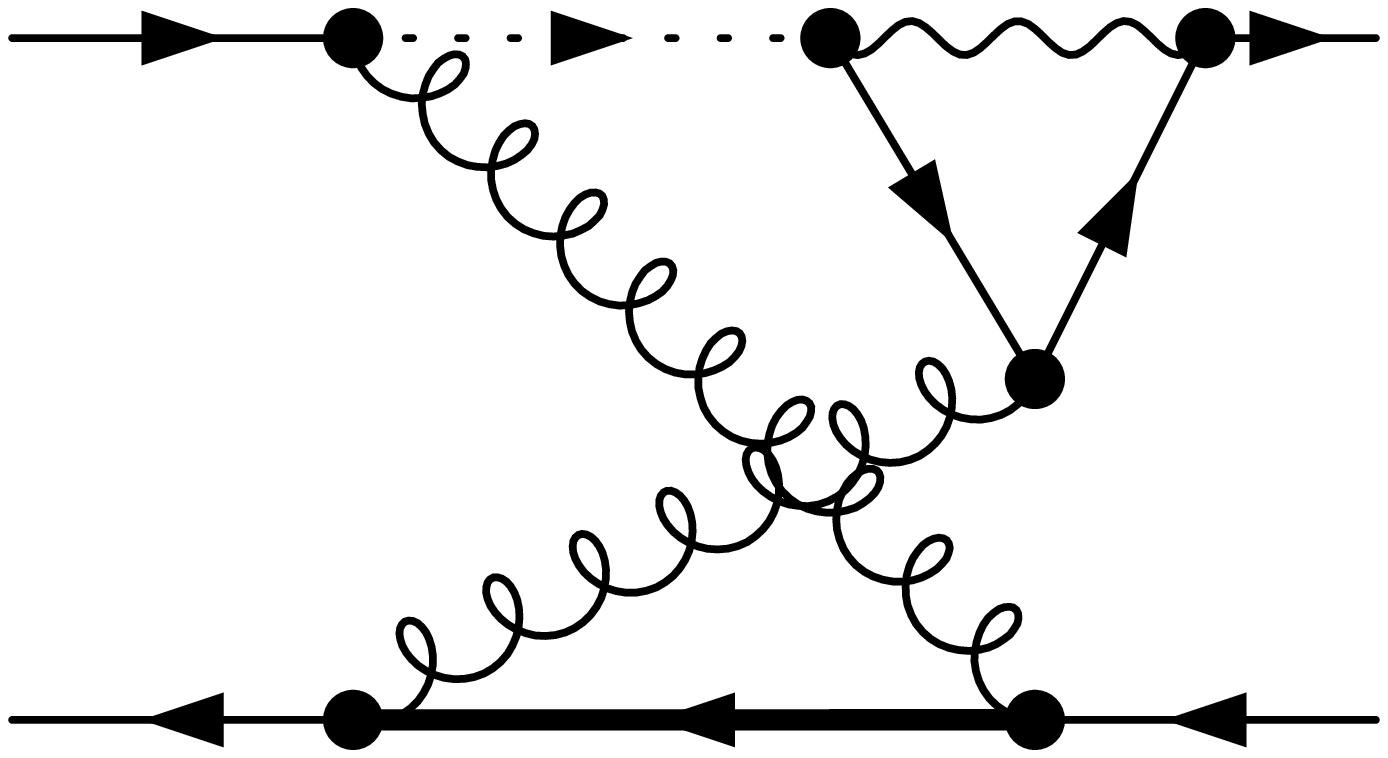}&
\includegraphics[width=3cm]{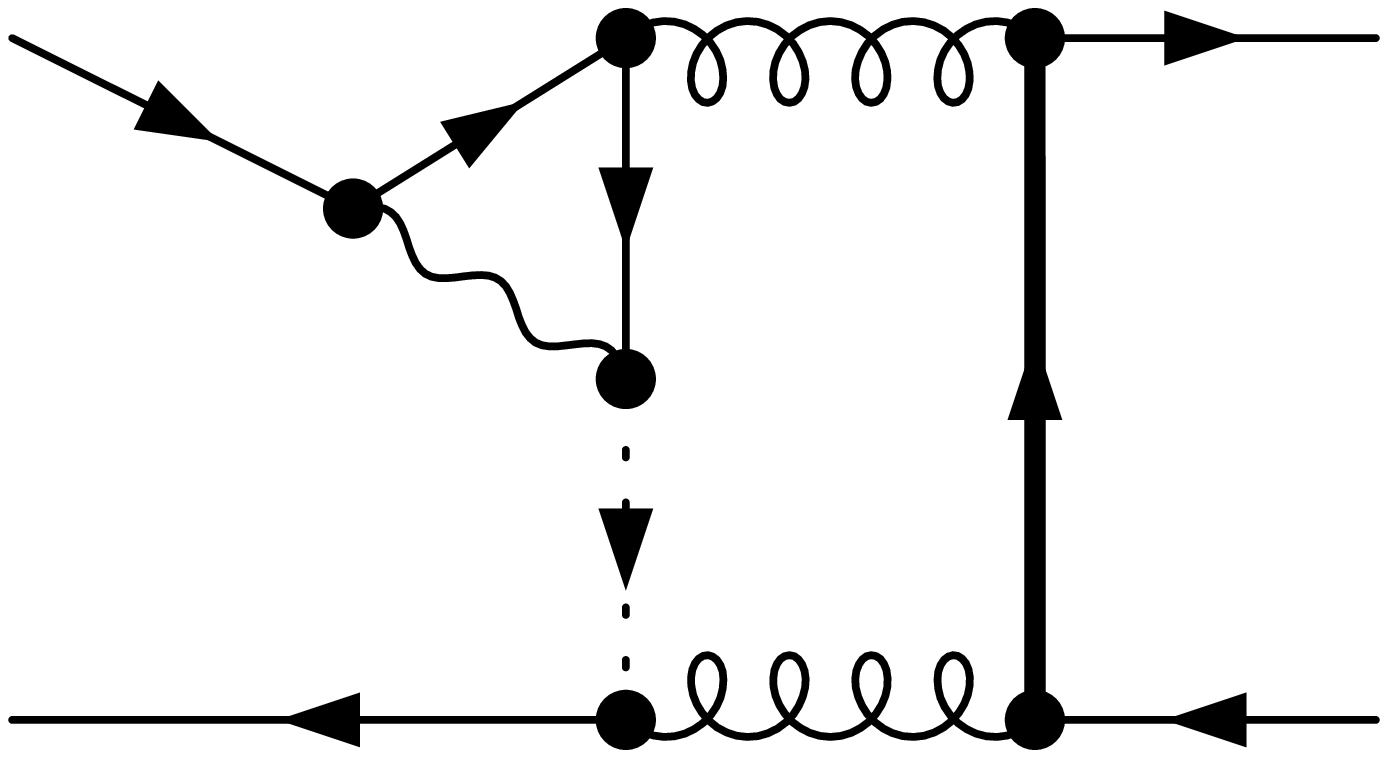}&
\includegraphics[width=3cm]{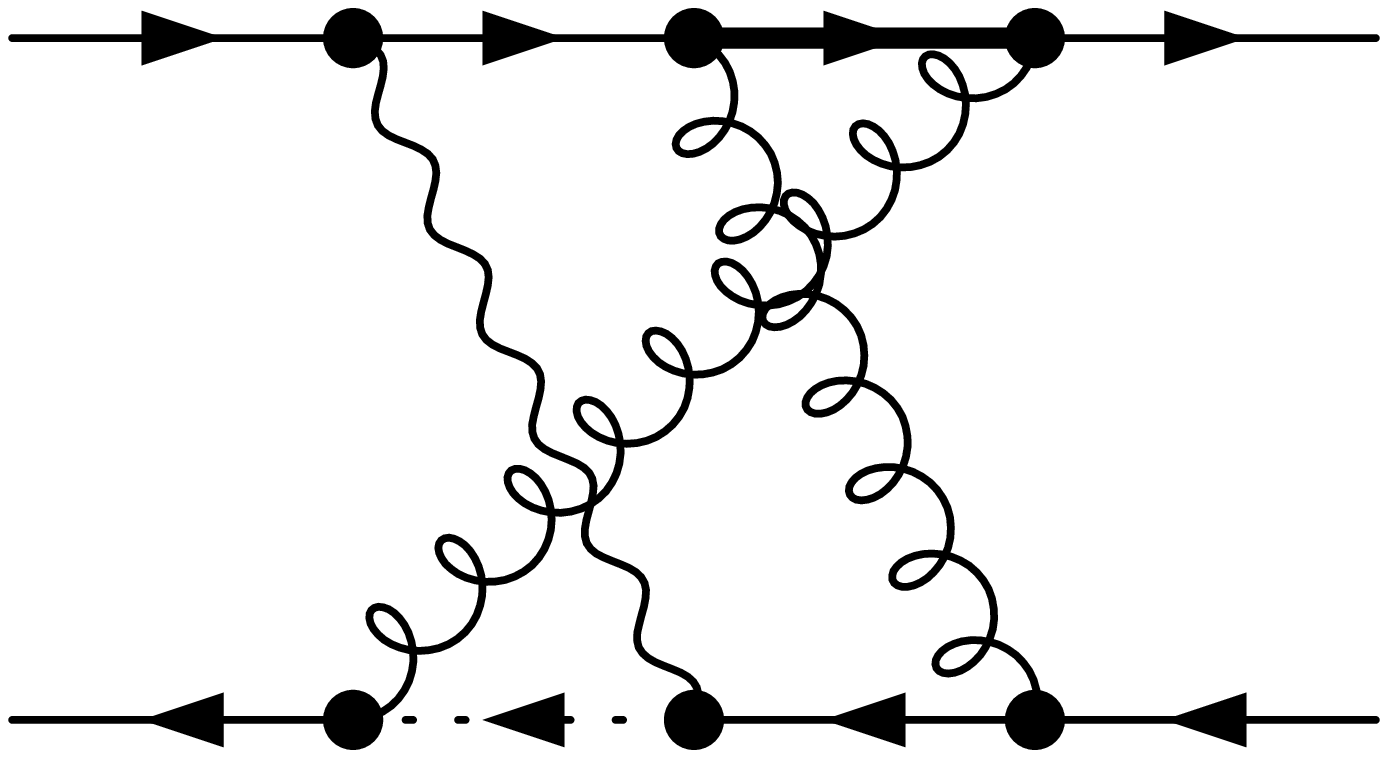}&
\includegraphics[width=3cm]{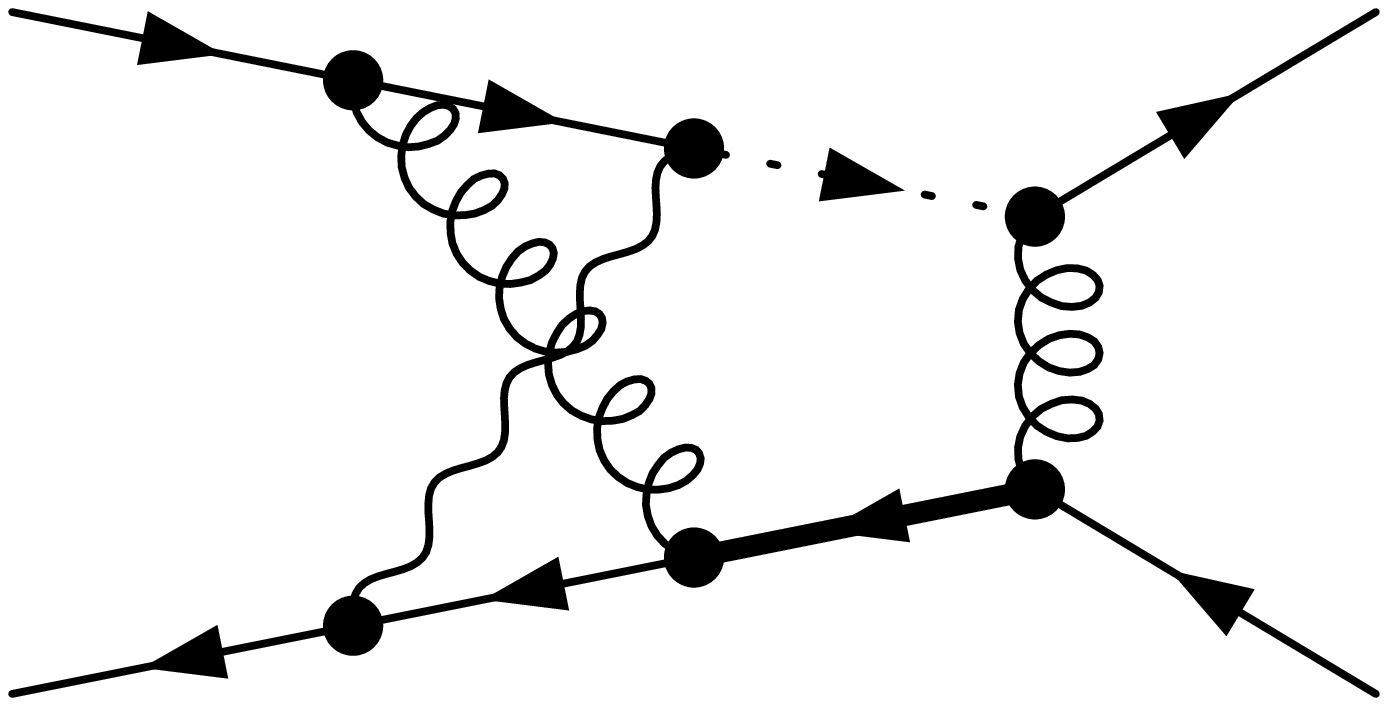}\\[-8mm]
(i)&(j)&(k)&(l)\\
\end{tabular}
\end{center}
\caption{\label{fig::2}  Two-loop  diagrams  with soft electron and soft  photon
exchange.  Dashed (thick) arrows correspond to the scalar soft (hard) electrons.
The loopy (wavy) lines correspond to  the eikonal  (soft) photons. Symmetric
diagrams are not shown. }
\end{figure}

The irreducible part  gets contributions from the Feynman diagrams with soft
electron pair exchange given in Fig.~\ref{fig::1}  and the Feynman diagrams with
both soft electron and soft photon exchanges, Fig.~\ref{fig::2}. Note that the
double-logarithmic corrections due to two soft photon exchanges cancel out in
the irreducible part according to the general factorization property of the
Sudakov logarithms. To compute the irreducible part we use the effective Feynman
rules described in the beginning of Sect.~\ref{sec::4}. The full set of
contributing diagrams is generated with {\textsc{Qgraf}~\cite{Nogueira:1991ex}}.
Its output is processed by a {\textsc{Mathematica}} program, which automatically
chooses the routing of internal and external momenta through the diagram in such
a way that the soft particle propagators carry only a single loop momentum and
no external momenta. The program generates {\textsc{FORM}}-readable expressions.
 By a  custom code written in
{\textsc{FORM}}~\cite{Vermaseren:2000nd,Kuipers:2012rf} the spin chains
appearing in the diagrams are projected into an irreducible basis, which allows
to easily square the amplitude.  The output is then  mapped into a set of five
two-loop ``master'' integrals $I_i$, which are evaluated  in the
double-logarithmic approximation in the Appendix. The soft electron pair
contribution is similar to the form factor corrections discussed in
\cite{Penin:2014msa} and can be reduced  to nonplanar and planar scalar vertex
integrals $I_{1,2}$. The  irreducible diagrams with soft photon exchange between
eikonal  lines, Figs.~\ref{fig::2}(a-d), are expressed through $I_2$ and the
product of the  one-loop integrals. The reduction of the diagrams
Figs.~\ref{fig::2}(e-h) includes the integral $I_3$, which depends on three
external momenta.  The diagrams with  soft  photon emission off the soft
electron line, Figs.~\ref{fig::2}(i,j) and Figs.~\ref{fig::2}(k,l), are reduced
to the  vector integrals $I_{4}$ and $I_{5}$, respectively.  The total
one-particle reducible contribution reads
\begin{eqnarray}
{{\rm d}\sigma^{(2)}_1\over{\rm d}\Omega}\Bigg|_{1PI}
&=&{1\over 3}{1+x^4\over (1-x)x^2}
-{21 - 70 x + 84 x^2 - 70 x^3 + 21 x^4\over 3(1-x)x^2}
\nonumber \\
&& -{34 - 184 x + 264 x^2 - 184 x^3 + 34 x^4 \over 3(1-x)x^2}
\,,
\label{eq::1pi}
\end{eqnarray}
where the three terms  correspond to the soft electron pair exchange, the soft
photon and soft electron exchange between the eikonal lines, and the soft photon
emission off the soft electron line, respectively.

The total result for the two-loop double-logarithmic power-suppressed term is
given by the sum of Eqs.~(\ref{eq::1lsquare}-\ref{eq::1pi}) and reads
\begin{eqnarray}
{{\rm d}\sigma^{(2)}_1\over{\rm d}\Omega}
&=&8{{\rm d}\sigma^{(0)}_1\over{\rm d}\Omega}
-{4 + 80 x - 360 x^2 + 476 x^3 - 360 x^4 + 80 x^5 + 4 x^6\over 3(1-x)x^3}
\nonumber\\
&=&-{4 + 176 x - 456 x^2 + 476 x^3 - 456 x^4 + 176 x^5 + 4 x^6
\over 3(1-x)x^3}\,.
\label{eq::result}
\end{eqnarray}
To estimate the numerical impact of the power-suppressed terms let us consider
the scattering  at  $\theta\sim 30^\circ$. For this scattering angle the
correction to the cross section is maximal at the energy $\sqrt{s}\approx 8\,
m_e\approx 4$~MeV, where
\begin{equation}
\rho\tau^2{{\rm d}\sigma^{(2)}_1\over{\rm d}\Omega}\approx
-24.6\left({\alpha\over\pi}\right)^2
{{\rm d}\sigma^{(0)}_0 \over{\rm d}\Omega}\,.
\label{eq::num}
\end{equation}
The effect decreases with the increasing energy but even for $\sqrt{s}\sim 300\,
m_e\approx 150$~MeV corresponding to $\rho\approx 10^{-5}$ the numerical
coefficient in Eq.~(\ref{eq::num}) is  approximately equal to  $-1$, {\it i.e.}
the  double logarithmic power-suppressed term is comparable to the
nonlogarithmic two-loop leading-power corrections.

\section{Summary}
\label{sec::summary}
In this paper we have developed a systematic approach  for the calculation of
the leading power correction to the high-energy  scattering processes in the
double logarithmic approximation. We focus on the  two-loop electron-positron
scattering in QED but the analysis can be extended to more complicated processes
and to nonabelian gauge theories. The higher order double-logarithmic
corrections in QED can in principle be resumed by using  the method described in
Refs.~\cite{Penin:2014msa,Melnikov:2016emg}. The general feature of the
high-energy expansion is the absence of the leading power-suppressed
double-logarithmic pure Sudakov corrections to the amplitudes due to the soft
virtual photon exchange. At the same time the structure of the corrections to
the two-particle scattering amplitudes turns out to be more diverse than for the
form factors describing single particle scattering in an external field. In
particular the non-Sudakov double logarithms appear already in  one-loop
scattering amplitude due to a single soft electron exchange. For the energies
ranging from a few to a few hundred MeV where $|\ln\rho|\gg 1$ and
$\rho\ln^4\rho\sim 1$,  the calculated two-loop double-logarithmic terms
saturate the power-suppressed contribution and are comparable in magnitude with
the two-loop nonlogarithmic leading-power corrections. This effectively sets up
the low boundary of the energy region  where the leading power approximation for
the ${\cal O}(\alpha^2)$ cross section can be used.

\section*{Acknowledgments}
We would like to thank  R. Bonciani,  T. Liu, K. Melnikov, and   V. Smirnov for
useful discussions and collaboration. We are grateful to M. Steinhauser for careful
reading the manuscript and useful comments. The work of A.P. is supported in
part by NSERC and Perimeter Institute of Theoretical Physics.


\appendix

\section{Evaluation of two-loop integrals}\label{AppendixA}
The one-particle irreducible diagrams in Fig.~\ref{fig::1} and  Fig.~\ref{fig::2}(a-d)
can be reduced to two scalar integrals
\begin{eqnarray}
I_1(p_i,p_j)&=&\int{d^4l_1}{d^4l_2}D(l_1)D(l_2)D(p_i+l_1+l_2)D(p_i+l_1)
D(p_j-l_1-l_2)D(p_j-l_2)\,,
\nonumber\\
&&\label{eq::I1}\\
I_2(p_i,p_j)&=&\int{d^4l_1}{d^4l_2}D(l_1)D(l_2)D(p_i+l_1+l_2)D(p_i+l_1)
D(p_j-l_1-l_2)D(p_j-l_1)\,,
\nonumber\\
&&\label{eq::I2}
\end{eqnarray}
where $D(k)=1/(k^2-m_e^2)$.  Let us consider the nonplanar case~(\ref{eq::I1}).
To compute the integral  in the double-logarithmic
approximation we follow Ref.~\cite{Sudakov:1954sw} and introduce the Sudakov
parametrization of each virtual momentum $l_k=u_kp_i+v_kp_j+{l_k}_\perp$.
Integration  over the transverse momentum components ${l_k}_\perp$ is performed by
taking the residue of a soft propagator pole
\begin{equation}
 D(l_k) \to -i\pi \delta(l_k^2 - m_e^2) =
- i \pi \delta(s u_kv_k - {l_k}_\perp^2 - m_e^2)\,.
\end{equation}
For $\rho<u_k,v_k<1$ the eikonal propagators become
\begin{eqnarray}
D(p_i +l_1)\approx \frac{1}{s_{ij}v_1},&&\qquad
D(p_i +l_1+l_2)\approx \frac{1}{s_{ij}(v_1+v_2)},
\nonumber\\
D(p_j -l_2)\approx -\frac{1}{s_{ij}u_2},&&\qquad
D(p_i -l_1-l_2)\approx -\frac{1}{s_{ij}(u_1+u_2)}\, .
\end{eqnarray}
Then the  double-logarithmic region is given by the interval $\rho<v_1<v_2<1$,
$\rho<u_2<u_1<1$ with an additional constraint  $\rho<u_kv_k$, which ensures
that the soft propagators can go on-shell. Thus in the double-logarithmic
approximation the two-loop nonplanar integral reads
\begin{equation}
I_1(p_i,p_j)\approx\left({i\pi^2 \over  s_{ij}}\right)^2
\int^1_{\rho}{{\rm d}v_1\over v_1}
\int^1_{v_1}{{\rm d}v_2\over v_2}
\int_{\rho/v_1}^1{{\rm d}u_1\over u_1}
\int_{\rho/v_2}^{u_1}{{\rm d}u_2\over u_2}\,.
\label{eq::I1sud}
\end{equation}
By introducing the normalized logarithmic variables $\eta_k=\ln
v_k/\ln\rho$ and $\xi_k=\ln u_k/\ln\rho$ Eq.~(\ref{eq::I1sud}) can be
transformed to
\begin{equation}
I_1(p_i,p_j)\approx N^2_{ij}
\int  \theta(1-\eta_1-\xi_1)\theta(1-\eta_2-\xi_2)
\theta(\eta_2-\eta_1) \theta(\xi_1-\xi_2)
 {\rm d}\eta_1{\rm d}\eta_2{\rm d}\xi_1{\rm d}\xi_2={N^2_{ij}\over 12}\,,
\label{eq::I1res}
\end{equation}
where $N_{ij}={i\pi^2 \ln^2\rho/  s_{ij}}$ and the integration goes over the
four-dimensional cube \linebreak $0<\eta_k,~\xi_k<1$.  The only difference in
calculation of the  planar two-loop integral~(\ref{eq::I2}) is  the ordering of
the variables  $\eta_2<\eta_1$, which provides the double-logarithmic scaling
of the integrand. Thus one gets
\begin{equation}
I_2(p_i,p_j)\approx N^2_{ij}
\int  \theta(1-\eta_1-\xi_1)\theta(1-\eta_2-\xi_2)
\theta(\eta_1-\eta_2) \theta(\xi_1-\xi_2)
 {\rm d}\eta_1{\rm d}\eta_2{\rm d}\xi_1{\rm d}\xi_2={N^2_{ij}\over 24}\,.
\label{eq::I2res}
\end{equation}
The diagrams where the soft photon and  soft electron emitted by the same
eikonal line end on different eikonal lines, Fig.~\ref{fig::2}(e-h), include the
scalar integral depending on three external momenta, which can be evaluated in
the same way
\begin{equation}
I_3(p_i,p_j,p_k)=
\int{d^4l_1}{d^4l_2}D(l_1)D(l_2)D(p_i+l_1)D(p_j-l_1)
D(p_j-l_1-l_2)D(p_k+l_2)\approx{N_{ij}N_{jk}\over 8}\,.
\label{eq::I3}\\
\end{equation}
The diagrams with the soft photon emission off the  soft electron line,
Fig.~\ref{fig::2}(i-l), include the electron propagator, which depend on both
soft momenta. When the  soft momenta are close to the mass shell the propagator
becomes eikonal
\begin{equation}
\frac{ \hat l_1 +\hat l_2 +m_e}{(l_1+l_2)^2 - m_e^2} \approx
\frac{\hat l_1 + \hat l_2}{2(l_1l_2)}
\end{equation}
and is sufficiently singular to produce the double-logarithmic contribution
despite  the presence of soft momenta in the numerator \cite{Melnikov:2016emg}.
In total we  have to take into account two vector master integrals, which depend
on two and three external momenta. It is convenient to project them on the
external momenta and consider the following quantities
\begin{eqnarray}
I_4(p_i,p_j)&=&
\int{d^4l_1}{d^4l_2}D(l_1)D(l_2)D(l_1-l_2)D(p_i+l_1)D(p_i+l_2)D(p_j-l_1)
\nonumber \\
&\times&
(l_1p_i)\,,
\label{eq::I4}\\
I_5(p_i,p_j,p_k)&=&
\int{d^4l_1}{d^4l_2}D(l_1)D(l_2)D(l_1+l_2)D(p_i-l_1)D(p_j-l_2)D(p_k+l_1+l_2)
\nonumber \\
&\times&
\left\{(l_1p_i),(l_1p_j),(l_1p_k)\right\}\,.
\label{eq::I5}
\end{eqnarray}
Let us consider the calculation of the integral~(\ref{eq::I4}). Only the 
case with  all the massive propagators is required. We introduce the
Sudakov parameters  in a slightly different way $l_1=u_1p_i+v_1p_j+{l_1}_\perp$,
$l_2=u_2p_i+v_2l_1+{l_2}_\perp$.  Then
\begin{equation}
 D(l_1 -l_2)\approx -\frac{1}{s_{ij}v_1u_2}
\end{equation}
and the extra factor $v_1$ in the denominator cancels the one from the scalar 
product $(l_1p_i)\approx s_{ij}v_1/2$ in the numerator providing the double 
logarithmic scaling of the integrand. The double-logarithmic integration region 
is now defined by the intervals $\rho<v_1,u_1<1$, $\rho/v_1<v_2,u_2<1$, 
$\rho<u_1v_1$, $\rho/v_1<u_2v_2$ and $\rho<v_1,u_1<1$, $\rho/v_1<v_2,u_2<1$, 
$\rho<u_1v_1$, $\rho/v_1<u_2v_2$, $u_2<u_1$,  which correspond to the 
contribution  of the poles of $D(l_1)$  and $D(l_1+l_2)$ propagators, 
respectively. These contributions are of the opposite sign so that one gets
\begin{eqnarray}
I_4(p_i,p_j)&\approx& s_{ij}N^2_{ij}
\int\theta(1-\eta_1-\xi_1)
\theta\left(1-\eta_1-\eta_2-\xi_2\right)
\theta(\xi_1-\xi_2)={N^2_{ij}s_{ij}\over 12}\,.
\label{eq::I4res}
\end{eqnarray}
In the same way we obtain 
\begin{equation}
I_5(p_i,p_j,p_k)|_{D(l_1)}\approx{\displaystyle
{1\over 24}\left\{{N_{jk}^2s_{jk}},{2N_{ik}^2s_{ik}},{2N_{ij}^2s_{ij}}\right\}}\,,
\label{eq::I5res}
\end{equation}
for the contribution of the pole of the $D(l_1)$ propagator to the Eq.~(\ref{eq::I5}). 
The contribution of the $D(l_1+l_2)$ pole can be easily obtained from this 
result by redefining  the external  momenta. In general for the three external 
momenta case we also need an infrared finite integral~(\ref{eq::I5}) with one 
massless propagator, which can be evaluated by the same technique.  
 

\end{document}